# Photonic Crystal Terahertz Leaky-Wave Antenna

Hichem Guerboukha[1], Jingwen Li[2]
[1] School of Engineering, Brown University, 02912, USA
[2] School of Science, Jiangnan University, 1800 Lihu Ave, Wuxi 214122, China

Corresponding author: hichem_guerboukha@brown.edu

**Abstract:**
Current wireless systems are experiencing unprecedented pressure that could be relieved by shifting carrier frequencies towards the terahertz spectrum. Yet, components are required to enable spectrally efficient communications at such high frequencies. Recently, the photonic crystal platform has been shown to enable signal processing at the chip-scale. Similarly, leaky-wave antennas were used to generate highly directional THz beams. Here, we combine photonic crystals with leaky-wave antennas. Practically, we propose to use an array of cylinders sandwiched between two plates, and we demonstrate a bandgap in the 250-350 GHz range by judicious geometric design. We introduce a linear defect in the crystal by removing a row of cylinders, allowing to guide the THz beam. Then, a careful study of the relation dispersion of the guided mode shows that its propagation constant is larger to that of free space. This allows the mode to efficiently outcouple when a thin aperture slot is introduced above the defect. We numerically demonstrate that the far-field diffracted beam can cover a wide angular range from 30 to 90 degrees. Finally, we show how the photonic crystals can be further engineered to enable coverage in two opposite angular directions.

## 1. Introduction

Current network systems are experiencing unprecedented pressure to keep up with the ever-increasing demand [1]. In the past couple of years, extensive research has been undertaken into the possibility of exploiting carrier frequencies in the terahertz band (0.1 – 10 THz) for wireless communications [2]. There, wide bandwidths allow ultra-high-speed data transmission in the order of Gbit/s [3–5]. However, working at such high frequencies brings its set of challenges. One of the issues is free-space path loss (FSPL) which can be as high as ~80 dB at 300 GHz for a 1-m distance. One of the strategies to counter these high losses is to use antennas that can generate highly directional beams, such as leaky-wave antennas.

Leaky-wave antennas have been thoroughly studied in the microwave range for many years now [6–9]. It is only recently that they have been proposed to address the challenges associated with THz communications, including FSPL [10]. The simplest leaky-wave antenna is made of a slot cut along the length of a waveguide. As the wave propagates at the slot interface, it emerges from the waveguide at a frequency-dependent angle. The produced rainbow has been demonstrated for many applications in the THz range, including beam steering [11,12], frequency-division multiplexing [13,14], link discovery [15], radar [16,17] and THz identification tag [18].

Furthermore, another crucial consideration when designing THz communication components is how to accommodate basic signal processes – e.g., filtering, modulation, and multiplexing as well as free space coupling – in an efficient and integrated manner. This is needed to achieve multilevel modulation schemes and multiplexing schemes that can maximize spectral efficiency. With these design perspective and paradigm in mind, great efforts have been applied to develop terahertz-wave platforms on which waves can be manipulated with sufficient confinement and efficient interaction. Up to date, several types of configurations based on different operation principles have been proposed and reported, some of these platforms include metamaterials [19], resonant cavities [20], as well as photonic crystal structures [21–23].

In particular, photonic crystals provide a versatile platform for wave manipulation and signal processing for THz communications. In fact, photonic crystals have been extensively studied as photonic components to guide and manipulate optical signals in light-wave regions over the last few decades. In general, photonic crystals are artificial periodic structures that control the propagation of light, similar to electrons being affected by ionic lattices. Through a judicious design of the geometry and material properties of their constituents, photonic bandgap may arise in the band diagram. One can then force light to propagate in any given direction by introducing a defect in the geometry. In the past several years, several types of THz photonic crystal waveguides have been proposed [24], including Bragg waveguides [25–27] and hyperuniform disordered waveguides [28] in the context of propagation [29,30] and sensing [31,32]. It is only recently that waveguides have been studied for application in communications [33,34]. However, most of these waveguides are designed to confine radiation throughout their length, which can be considered as unidimensional dimensional THz communication link. To develop compact components for diverse THz manipulations – e.g., multiplexing and routing of terahertz waves – photonic crystal waveguides can be implemented in the planar form (i.e., 2D photonic slab). For example, Withayachumnankul et al [aa] reported a photonic slab with a hexagonal array of air holes for the efficient guiding and manipulation of THz waves [35]. By introducing a line defect in the periodicity, a waveguiding structure is created. The clad region around the defect region provides the bandgap effect and results in wave confinement in the in-plane dimensions. Meanwhile, total internal reflection prevents radiation leakage into free space in the vertical direction. With such configuration, frequency-division triplexer has been realized on the photonic crystal platform experimentally.

Although various configurations of 2D photonic crystal slabs have been reported in the terahertz-wave region [23,36–38] for different manipulations, none of these studies focus on a detailed design of antennas based on photonic crystal slabs. Antennas are vital for coupling the guided waves in the photonic crystal to the free space waves, which is of significant importance in order to realize THz wireless communications.

In this work, we propose to use a photonic crystal platform for the development of THz leaky-wave antennas. Here, we demonstrate how one can design a photonic crystal waveguide to leak radiation in a controlled manner. We begin by designing a photonic crystal made of an array of alumina cylinders sandwiched between two metallic plates. Then, we introduce a linear defect by removing a row of cylinders, and we show how the wave propagates in the defect. A closer look at the relation dispersion of the defect mode reveals that its wavevector is located in the radiation region, that is $\beta > \omega/c$. This realization then allows us to efficiently leak the energy out of the waveguide by the cutting a thin slot on the metallic plate right above the linear defect. We then show how the radiation angle evolves as a function of the frequency.

Fig. 1a depicts the schematic of the proposed photonic crystal. It is made of an array of alumina cylinders (refractive index of 3.05 [39]) embedded in air and distributed on a square lattice with a lattice constant of 388 μm. The cylinders have diameters of 120 μm and heights of 300 μm. They are sandwiched between two metallic plates. Fig. 1b shows the calculated band diagram of the resulting photonic crystal. As it can be seen, this geometry opens a bandgap at 300 GHz with a bandwidth of ~95 GHz ($\Delta\nu/\nu_0 = 32\%$). This means that a device made with this geometry would essentially act as a filter and block the transmission of waves located in the bandgap. This is shown in Fig. 1c where the crystal is excited with a planar wave of frequencies of 240 GHz, 300 GHz and 380 GHz, respectively below, inside and above the bandgap. As can be seen, the wave located in the bandgap cannot propagate, while it propagates when its frequency is located either below or above the bandgap.

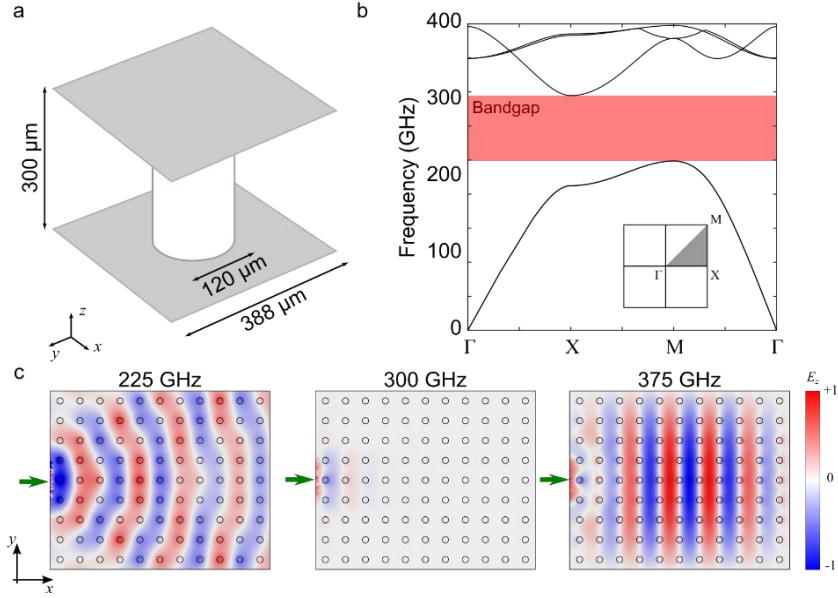

**Fig. 1.** (a) Schematic of the photonic crystal unit cell. (b) Band diagram of the photonic crystal. The bandgap is shown in red. (c) Propagation of waves in the photonic crystal for 3 frequencies below, inside and above the bandgap, respectively 240 GHz, 300 GHz and 370 GHz.

Next, we introduce a linear defect by removing one row of cylinders in the photonic crystal. Fig. 2a shows the resulting relation dispersion for $k_x$, i.e. in the $\Gamma \rightarrow X$ direction. As can be seen, a unique defect mode exists in the bandgap (green curve in Fig. 2a). Fig. 2b shows that the electric field distribution of the defect mode is polarized in $z$ (parallel to the cylinders). As expected, it is mainly localized in the defect (the missing row). Therefore, when THz waves of vertical polarization are injected at the input port, they will follow the defect in the crystal. This is shown in Fig. 2c for linear and L-shaped defects.

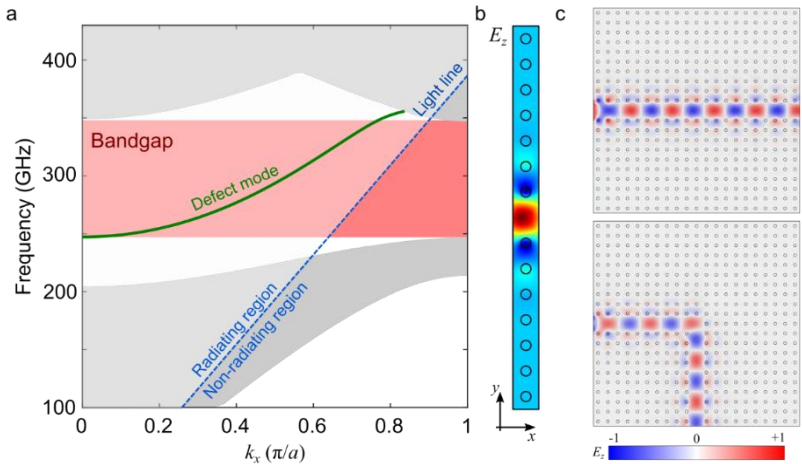

**Fig. 2.** (a) Dispersion relation in the $\Gamma \rightarrow X$ direction when a defect is introduced in the photonic crystal. The green curve is the defect mode located in the bandgap (red box). The dotted line is the light line $k_x = \omega/c$ that delimits the radiating region (top) from the non-radiating region (bottom). (b) Distribution of the electric field amplitude for the defect mode. Most of the field is located in the defect. (c) Examples of waveguiding using a linear straight defect (top) and a L-shaped defect (bottom).

From the relation dispersion of Fig. 2a, we note that the defect mode is in the radiation region i.e., its wavevector is above the light line and follows the inequality $\beta > \omega/c$. This means that this mode can directly outcouple in free space by phase matching. This can be done by simply introducing a thin linear slot in the metal plate right above the defect. This is shown in Fig. 3a at 300 GHz for a slot width of 0.2 mm ($\lambda/1.5$). There, the radiation peaks at an angle $\theta$ defined with $\theta = \mathrm{acos}(\beta/k_0)$, where $k_0$ is the free space wavevector, meaning that $\theta$ is also a function of the frequency. The radiation pattern is shown in Fig. 3b. As expected, the numerically calculated radiation pattern follows well the angle that can be derived from the phase matching condition (white curve in Fig. 3b).

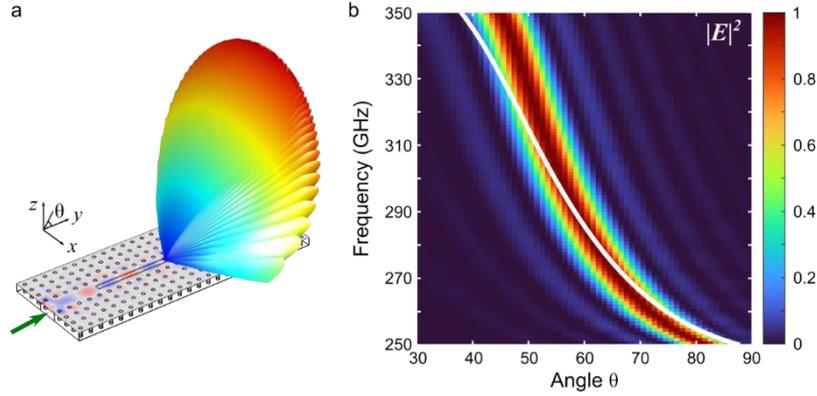

**Fig. 3.** (a) 3D far-field radiation pattern (power) when a slot is introduced in the metal plate right above the linear defect. (b) The radiation pattern as a function of frequency, and (b) for different frequencies.

Finally, we demonstrate the ability of the photonic crystal platform to engineer the far-field radiation pattern. For that purpose, we design a photonic crystal where cylinders have been removed to create a T-junction. The resulting electric field after exciting the waveguide with a vertical polarization is shown in Fig. 4a and follows the shape of the defect. Then, two slots are cut above the T-junction. Because the wave propagates in counter directions inside the waveguide, it outcouples at two different angles $\pm \theta$, also given by the phase matching condition (Fig. 4b).

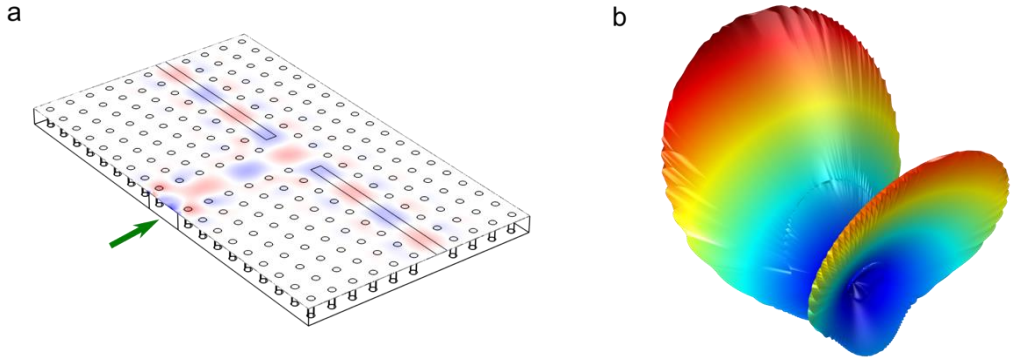

**Fig. 4.** (a) Electric field distribution in the T-junction after exciting the photonic crystal with a vertically polarized beam. (b) Resulting far-field pattern (power).

**Conclusion**

In conclusion, we demonstrated a photonic crystal leaky-wave antenna operating at THz frequencies. The photonic crystal was designed with an array of alumina cylinders sandwiched between two metallic plates. A defect was introduced by removing a row of cylinders, effectively creating a path for light to propagate. The resulting defect mode being located in the radiation region, the simple introduction of a thin slot on the metallic plate right above the linear defect allows to efficiently outcouple the wave in free space. We showed that the radiation pattern peaked at the frequency given by the phase matching condition between the propagating wave and the free space wavevector. Finally, we showed the versatility of the photonic crystal platform by creating a T-junction, which allowed to output the wave in two opposite directions. This work is a first step towards the effective engineering of THz far-field radiation patterns using the photonic crystal platform.

## References

1. Cisco, "Cisco Annual Internet Report (2018–2023) White Paper (March 2020 Update)," https://www.cisco.com/c/en/us/solutions/collateral/executive-perspectives/annual-internet-report/white-paper-c11-741490.html.
2. T. Nagatsuma, G. Ducournau, and C. C. Renaud, "Advances in terahertz communications accelerated by photonics," Nat. Photonics **10**, 371–379 (2016).
3. S. Koenig, D. Lopez-Diaz, J. Antes, F. Boes, R. Henneberger, A. Leuther, A. Tessmann, R. Schmogrow, D. Hillerkuss, R. Palmer, T. Zwick, C. Koos, W. Freude, O. Ambacher, J. Leuthold, and I. Kallfass, "Wireless sub-THz communication system with high data rate," Nat. Photonics **7**, 977–981 (2013).
4. T. Harter, C. Füllner, J. N. Kemal, S. Ummethala, J. L. Steinmann, M. Brosi, J. L. Hesler, E. Bründermann, A. S. Müller, W. Freude, S. Randel, and C. Koos, "Generalized Kramers–Kronig receiver for coherent terahertz communications," Nat. Photonics (2020).
5. K. Nallappan, H. Guerboukha, C. Nerguizian, and M. Skorobogatiy, "Live Streaming of Uncompressed HD and 4K Videos Using Terahertz Wireless Links," IEEE Access **6**, 58030–58042 (2018).
6. L. Goldstone and A. Oliner, "Leaky-wave antennas I: Rectangular waveguides," IRE Trans. Antennas Propag. **7**, 307–319 (1959).
7. A. Oliner and D. R. Jackson, "Leaky-Wave Antennas," in *Antenna Engineering Handbook* (2007), p. 12.
8. D. R. Jackson, C. Caloz, and T. Itoh, "Leaky-Wave Antennas," Proc. IEEE **100**, 2194–2206 (2012).
9. L. Liu, C. Caloz, and T. Itoh, "Dominant mode leaky-wave antenna with backfire-to-endfire scanning capability," Electron. Lett. **38**, 1414–1416 (2002).
10. H. Guerboukha, R. Shrestha, J. Neronha, O. Ryan, M. Hornbuckle, Z. Fang, and D. M. Mittleman, "Efficient leaky-wave antennas at terahertz frequencies generating highly directional beams," Appl. Phys. Lett. **117**, 261103 (2020).
11. M. Esquius-Morote, J. S. Gomez-Diaz, and J. Perruisseau-Carrier, "Sinusoidally modulated graphene leaky-wave antenna for electronic beamscanning at THz," IEEE Trans. Terahertz Sci. Technol. **4**, 116–122 (2014).
12. X. C. Wang, W. S. Zhao, J. Hu, and W. Y. Yin, "Reconfigurable terahertz leaky-wave antenna using graphene-based high-impedance surface," IEEE Trans. Nanotechnol. **14**, 62–69 (2015).
13. N. J. Karl, R. W. McKinney, Y. Monnai, R. Mendis, and D. M. Mittleman, "Frequency-division multiplexing in the terahertz range using a leaky-wave antenna," Nat. Photonics **9**, 717–720 (2015).
14. J. Ma, N. J. Karl, S. Bretin, G. Ducournau, and D. M. Mittleman, "Frequency-division multiplexer and demultiplexer for terahertz wireless links," Nat. Commun. **8**, (2017).
15. Y. Ghasempour, R. Shrestha, A. Charous, E. Knightly, and D. M. Mittleman, "Single-shot link discovery for terahertz wireless networks," Nat. Commun. **11**, (2020).
16. K. Murano, I. Watanabe, A. Kasamatsu, S. Suzuki, M. Asada, W. Withayachumnankul, T. Tanaka, and Y. Monnai, "Low-Profile Terahertz Radar Based on Broadband Leaky-Wave Beam Steering," IEEE Trans. Terahertz Sci. Technol. **7**, 60–69 (2017).
17. Y. Amarasinghe, R. Mendis, and D. M. Mittleman, "Real-time object tracking using a leaky THz waveguide," Opt. Express **28**, 17997 (2020).
18. Y. Amarasinghe, H. Guerboukha, Y. Shiri, and D. M. Mittleman, "Bar code reader for the THz region," Opt. Express **29**, 20240 (2021).
19. E. Herrmann, H. Gao, Z. Huang, S. R. Sitaram, K. Ma, and X. Wang, "Modulators for mid-infrared and terahertz light," J. Appl. Phys. **128**, 140903 (2020).
20. Y. Akahane, T. Asano, B.-S. Song, and S. Noda, "High-Q photonic nanocavity in a two-dimensional photonic crystal," Nature **425**, 944–947 (2003).
21. S. Noda, A. Chutinan, and M. Imada, "Trapping and emission of photons by a single defect in a

photonic bandgap structure," Nature **407**, 608–610 (2000).
22. M. Thorhauge, L. H. Frandsen, and P. I. Borel, "Efficient photonic crystal directional couplers," Opt. Lett. **28**, 1525 (2003).
23. K. Tsuruda, M. Fujita, and T. Nagatsuma, "Extremely low-loss terahertz waveguide based on silicon photonic-crystal slab," Opt. Express **23**, 31977 (2015).
24. K. Nallappan, H. Guerboukha, Y. Cao, G. Xu, C. Nerguizian, D. M. Mittleman, and M. Skorobogatiy, "Terahertz waveguides for next generation communication networks: needs, challenges and perspectives," in *Next Generation Wireless Terahertz Communication Networks*, and A. D. S. Ghafoor, M. H. Rehmani, ed. (CRC Press, Taylor and Francis Group, 2021).
25. J. Li, K. Nallappan, H. Guerboukha, and M. Skorobogatiy, "3D printed hollow core terahertz Bragg waveguides with defect layers for surface sensing applications," Opt. Express **25**, 4126 (2017).
26. T. Ma, K. Nallapan, H. Guerboukha, and M. Skorobogatiy, "Analog signal processing in the terahertz communication links using waveguide Bragg gratings: example of dispersion compensation," Opt. Express **25**, 11009 (2017).
27. Y. Cao, K. Nallappan, H. Guerboukha, T. Gervais, and M. Skorobogatiy, "Additive manufacturing of resonant fluidic sensors based on photonic bandgap waveguides for terahertz applications," Opt. Express **27**, 27663 (2019).
28. T. Ma, H. Guerboukha, and M. Skorobogatiy, "3D printed hollow-core terahertz optical waveguides with hyperuniform disordered dielectric reflectors," 2017 Conf. Lasers Electro-Optics, CLEO 2017 - Proc. **2017-Janua**, 1–2 (2017).
29. A. Markov, H. Guerboukha, and M. Skorobogatiy, "Hybrid metal wire–dielectric terahertz waveguides: challenges and opportunities [Invited]," J. Opt. Soc. Am. B **31**, 2587 (2014).
30. S. Atakaramians, S. Afshar V., T. M. Monro, and D. Abbott, "Terahertz dielectric waveguides," Adv. Opt. Photonics **5**, 169 (2013).
31. J. Li, H. Qu, and J. Wang, "Photonic Bragg waveguide platform for multichannel resonant sensing applications in the THz range," Biomed. Opt. Express **11**, 2476 (2020).
32. H. Guerboukha, G. Yan, O. Skorobogata, and M. Skorobogatiy, "Silk foam terahertz waveguides," Adv. Opt. Mater. **2**, 1181–1192 (2014).
33. K. Nallappan, Y. Cao, G. Xu, H. Guerboukha, C. Nerguizian, and M. Skorobogatiy, "Dispersion-limited versus power-limited terahertz communication links using solid core subwavelength dielectric fibers," Photonics Res. **8**, 1757 (2020).
34. Y. Cao, K. Nallappan, H. Guerboukha, G. Xu, and M. Skorobogatiy, "Additive manufacturing of highly reconfigurable plasmonic circuits for terahertz communications," Optica **7**, 1112 (2020).
35. W. Withayachumnankul, M. Fujita, and T. Nagatsuma, "Integrated Silicon Photonic Crystals Toward Terahertz Communications," Adv. Opt. Mater. **6**, 1800401 (2018).
36. M. Yata, M. Fujita, and T. Nagatsuma, "Photonic-crystal diplexers for terahertz-wave applications," Opt. Express **24**, 7835 (2016).
37. R. Kakimi, M. Fujita, M. Nagai, M. Ashida, and T. Nagatsuma, "Capture of a terahertz wave in a photonic-crystal slab," Nat. Photonics **8**, 657–663 (2014).
38. P. Gopalan and B. Sensale-Rodriguez, "2D Materials for Terahertz Modulation," Adv. Opt. Mater. **8**, 1900550 (2020).
39. J. Ornik, M. Sakaki, M. Koch, J. C. Balzer, and N. Benson, "3D Printed Al 2 O 3 for Terahertz Technology," IEEE Access **9**, 5986–5993 (2021).